# Spectrum of coupled waves in orthorhombic multiferroics with cycloidal antiferromagnetic structure in external electric and magnetic fields

Igor V. Bychkov, Dmitry A. Kuzmin, and Vladimir G. Shavrov

*Abstract*—Spectrum of coupled spin, acoustic, electro-dipole and electromagnetic waves in orthorhombic perovskite multiferroics placed in external electric and magnetic field has been calculated. Frequency and field dependencies of electric, magnetic, and magnetoelectric susceptibilities of multiferroics with cycloidal antiferromagnetic structure in external electric and magnetic fields have been investigated as well. The spectrum of coupled waves has a band structure. Width and frequency ranges of bands depend on external fields. The possibility of control of electrodynamic properties of orthorhombic perovskite multiferroics with cycloidal antiferromagnetic structure by external electric and magnetic fields has been shown.

## I. Introduction

Nowadays, multiferroics – materials with magnetic, electric, and often also elastic ordering, attract considerable attention for their non-trivial physical properties. Mainly, the recent interest is triggered by the discovery of the cross-coupling between magnetic and electric ordering in orthorhombically distorted perovskite manganites, RMnO$_3$ (R = Gd, Tb, Dy) [1, 2]. In such materials spontaneous polarization is caused by cycloidal-modulated spin order. Orthorhombic perovskite multiferroic TbMnO$_3$ ($P_{bnm}$), for example, has a cycloidal antiferromagnetic structure at temperatures T < 28 K [3]. The modulated magnetic structure itself contributes a number of features in the electro-dynamical properties of material [4, 5]. The coupled spin, electro-dipole and electromagnetic waves in TbMnO$_3$ with cycloidal antiferromagnetic structure, it's susceptibility tensors and dynamics of electromagnons in manganites had been theoretically investigated earlier [6-11], however the influence of external electric and magnetic fields on spectrum of coupled spin, electro-dipole, acoustic and electromagnetic waves in such material are not studied enough. The present work is devoted to studying of electrodynamical properties of orthorhombic multiferroics placed in an external electric and magnetic fields of different directions. Investigation of dynamics of the coupled excitations in the modulated magnetic structures are carried out in approach $L \gg b$, where $L = 2\pi/k$ is the period of modulated structure, $b$ is the lattice constant, $k$ is the wave number of modulated structure, when the phenomenological method is applicable. For example, in TbMnO$_3$, $k \approx 0.28b^{-1}$ [3], e.g. $L \approx 22b$, in GdMnO$_3$, $k \approx 0.2b^{-1}$ ($L \approx 31b$) [12], in DyMnO$_3$, $k \approx 0.35b^{-1}$ ($L \approx 18b$) [12] and phenomenological method can be used.

## II. Ginzburg-Landau Functional of Orthorhombic Multiferroics

Method of Lagrange has been used for investigating of dynamics of orthorhombic multiferroic. The Lagrangian's expression is $L = E - F$, where $E$ - the kinetic energy, $F$ - the Ginzburg-Landau function. In case of orthorhombic multiferroic:

$$E = V^{-1} \int 0.5 \left( \mu \dot{\mathbf{A}}^2 + \tilde{\mu} \dot{\mathbf{M}} + \lambda \dot{\mathbf{P}}^2 + \rho \dot{\mathbf{u}}^2 \right) d\mathbf{r};$$

$$F = \frac{1}{V} \int \{ \frac{a}{2}\mathbf{A}^2 + \frac{w}{2}A_z^2 + \frac{u}{4}\mathbf{A}^4 + \frac{\gamma}{2}(\partial_y \mathbf{A})^2 + \frac{\alpha}{2}(\partial_y^2 \mathbf{A})^2 -$$
$$-\mathbf{MH} + \frac{\beta}{2}\mathbf{M}^2 + \frac{\lambda_1}{2}(\mathbf{AM})^2 + \frac{\lambda_2}{2}\mathbf{A}^2\mathbf{M}^2 + \frac{b}{2}\mathbf{P}^2 - \qquad (1)$$
$$-\mathbf{PE} + \nu P_z \left( A_z \partial_y A_y - A_y \partial_y A_z \right) +$$
$$+ c_{ijlm} u_{ij} u_{lm} + b_{ijlm} A_i A_j u_{lm} + d_{ijlm} P_i P_j u_{lm} \} d\mathbf{r}.$$

Where $\mathbf{M} = \mathbf{M_1} + \mathbf{M_2} + \mathbf{M_3} + \mathbf{M_4}$ is the magnetization of the crystal; $\mathbf{A} = \mathbf{M_1} - \mathbf{M_2} - \mathbf{M_3} + \mathbf{M_4}$ is the vector of antiferromagnetism; $\mathbf{M}_i$ is a magnetization of sub lattices; $\mathbf{P}$ is the vector of polarization; $\mathbf{P}$ is the vector of polarization; $u_{ij} = \left( \partial u_i / \partial x_j + \partial u_j / \partial x_i \right)/2$ is the tensor of deformations; $\mathbf{u}$ is the displacement vector; $w$ is the constants of anisotropy; $a, u, \beta, \lambda_1, \lambda_2$ are the constants of homogeneous exchange; $\alpha, \gamma$ are the constants of heterogeneous exchange; $b, \nu$ are the constants of electric and magnetoelectric interactions; $c_{ijlk}, b_{ijlk}, d_{ijlk}$ are the tensors of elasticity, magnetostriction and electrostriction; $\lambda = m v_c / z^2$, where $z$ and $m$ are the charge and the reduced mass of the elementary cell with the

This work was supported in part by grants of RFBR №11-02-12255, №13-07-00462

I. V. Bychkov is with the Chelyabinsk State University, Chelyabinsk 454001 Russia (e-mail: bychkov@csu.ru).

D. A. Kuzmin is with the Chelyabinsk State University, Chelyabinsk 454001 Russia (e-mail: kuzmin@csu.ru).

V. G. Shavrov is with The Institute of Radioengineering and Electronics of RAS, Moscow 125009 Russia (e-mail: shavrov@mail.cplire.ru).

volume $v_c$; $\mu = \chi_\perp / 8g^2 M_0^2$, where $\chi_\perp$ is the static transversal magnetic susceptibility, $g$ is the gyromagnetic ratio, $M_0$ is the magnetization of saturation. In (1) we take into account only the biggest terms. The physical properties of the related terms have been studied in detail in [13, 14].

The ground non-collinear state is described by vectors of antiferromagnetism and polarization with following components: $P_{0x} = P_{0y} = 0$; $P_{0z} = P_0$; $A_{0x} = 0$; $A_{0y} = A_1 \sin ky$; $A_{0z} = A_2 \cos ky$. Here, $y$-axis is the modulation axis; $z$-axis is the direction of the spontaneous polarization.

Expressions for determining the parameters of the ground state may be obtained from the minimum of Ginzburg-Landau function. In case of external electric field $\mathbf{E}$ is collinear to a spontaneous polarization $\mathbf{P_{0z}}$ and external magnetic field have all non-zero components:

$$M_{0i}\beta_i = H_{0i}, \quad i = x, y, z;$$
$$\alpha k^3 \left( A_1^2 + A_2^2 \right) + 0.5\gamma k \left( A_1^2 + A_2^2 \right) = \nu A_1 A_2 P_0;$$
$$p_3 P_0^3 + P_0 \left( p_{1a1} A_1^2 + p_{1a2} A_2^2 + b \right) = E_{0z} + \nu k A_1 A_2; \quad (2)$$
$$a_{1,23} A_{1,2}^3 + A_{1,2} \begin{pmatrix} a_{1,21p} P_0^2 + a_{1a2} A_2^2 + a_{1,21my} \mathbf{M}_0^2 + \\ +4\left( \alpha k^4 + \gamma k^2 + a + w\delta_{21,2} \right) \end{pmatrix} = $$
$$= 8\nu k P_0 A_{2,1}.$$

Here, the following notation has been introduced:
$\beta_x = \left[ \lambda_2 \left( A_1^2 + A_2^2 \right)/2 + \beta \right]$, $\beta_{y,z} = \left[ \tilde{\lambda} A_{1,2}^2/2 + \lambda_2 A_{2,1}^2/2 + \beta \right]$,
$\tilde{\lambda} = \lambda_1 + \lambda_2$, $\Delta = \det |c_{ij}|$, $p_3 = -8\Delta^{-1}(\mathbf{D_3}\Delta_{\mathbf{D3}})$,
$p_{1a1,2} = -4\Delta^{-1}(\mathbf{B_{2,3}}\Delta_{\mathbf{D3}})$, $a_{1,23} = 3u - 24\Delta^{-1}(\mathbf{B_{2,3}}\Delta_{\mathbf{B2,3}})$,
$a_{1,21p} = -32\Delta^{-1}(\mathbf{D_3}\Delta_{\mathbf{B2,3}})$, $a_{11my} = 4\tilde{\lambda}$, $a_{21my} = 4\lambda_2$,
$a_{1a2} = u - B_{44}^2 c_{44}^{-1} - 8\Delta^{-1}(\mathbf{B_3}\Delta_{\mathbf{B2}})$, $\Delta_{\Xi i} = \left( \Delta_{\Xi i}^1, \Delta_{\Xi i}^2, \Delta_{\Xi i}^3 \right)$,
$\Xi_i = (\Xi_{i1}, \Xi_{i2}, \Xi_{i3})$, $\Xi = B, D$, $i = 1, 2, 3$, $\Delta_{\Xi i}^j$ can be obtained from expression for $\Delta$ by changing of $j$ column to vector $\Xi_i$, $\delta_{ij}$ is the Cronecker's symbol.

### III. ELECTRO-DYNAMICAL CHARACTERISTICS

For investigation of dynamic characteristics one have to take into account the system of Lagrange equations for $\mathbf{A}$, $\mathbf{M}$, $\mathbf{P}$ and $\mathbf{u}$. Solving the system of equations by method of low oscillations, linearizing and using the form of harmonic series for variables, in approach of first harmonics for the waves, propagating along $y$-axis, the oscillating amplitudes of polarization $\mathbf{p}$ and antiferromagnetism $\mathbf{a}$ vectors can be expressed as $p_i = \alpha_{ij} e_j + \kappa_{ij}^{em} h_j$, $m_i = \chi_{ij} h_j + \kappa_{ij}^{me} e_j$, where $\alpha_{ij}$, $\chi_{ij}$ are the tensors of electric and magnetic susceptibility, consequently, $\kappa_{ij}^{me} = \left( \kappa_{ij}^{em} \right)^*$ is the magnetoelectric susceptibility tensor.

The expressions for susceptibilities tensors on example of TbMnO$_3$ placed in external magnetic and electric fields have been obtained in [11]. Electric and magnetic susceptibility tensors are diagonals. Magnetoelectric susceptibility tensor has different non-zero components, depending on the direction of external magnetic field. For example, $\mathbf{H} \| \mathbf{x}$ activates only one component $\kappa_{xz}^{me}$; $\mathbf{H} \| \mathbf{y}$ and $\mathbf{H} \| \mathbf{z}$ activates two components $\kappa_{yz}^{me}$, $\kappa_{zy}^{me}$ and $\kappa_{yy}^{me}$, $\kappa_{zz}^{me}$, consequently.

Let us examine the details of the one of cases, when $\mathbf{H} \| \mathbf{x}$. Tensors $\alpha_{ij}$ and $\chi_{ij}$ have the following components:

$$\alpha_{xx} = \lambda^{-1} \left( \omega_{px}^2 - \omega^2 \right)^{-1}; \quad \chi_{xx} = \tilde{\mu}^{-1} \lambda \alpha_{zz} \Delta_{pz} \Delta_{mx}^{-1} \left( \omega^2 - \omega_l^2 \right)^{-1};$$

$$\alpha_{zz} = \frac{\Delta_{mx} \left( \omega^2 - \omega_l^2 \right) \left( \Delta_{ay}^+ \Delta_{az}^- + \left[ \Delta_{ay}^{+az} \right]^2 \right)}{\lambda \left( \Delta_{mx} \Delta_{pz} - \Delta_{mx}^{pz} \Delta_{pz}^{mx} \right)};$$

$$\chi_{yy} = \tilde{\mu}^{-1} \Delta_{ax}^+ \left\{ \left( \omega^2 - \omega_{my}^2 \right) \Delta_{ax}^+ + \omega_{1xy}^2 \left( \omega^2 - \omega_{tx}^2 \right)/2 \right\}^{-1};$$

$$\chi_{zz} = \tilde{\mu}^{-1} \left( \omega^2 - \omega_{ax}^{-\,2} \right) \left\{ \omega_{1xz}^2 \omega_{2xz}^2 / 2 - \left( \omega^2 - \omega_{mz}^2 \right) \left( \omega^2 - \omega_{ax}^{-\,2} \right) \right\}^{-1};$$

$$\alpha_{yy} = \lambda^{-1} \left( \omega^2 - \omega_{tz}^2 \right) \left( \Delta_{ay}^- \Delta_{az}^+ + \left[ \Delta_{az}^{+ay} \right]^2 \right) \times$$
$$\times \left\{ \begin{aligned} & \Delta_{py} \left( \Delta_{ay}^- \Delta_{az}^+ + \left[ \Delta_{az}^{+ay} \right]^2 \right) + \\ & + \omega_{D44}^z \omega_l^2 \omega_{tz}^2 \begin{bmatrix} i\omega_{B44}^{zy\,2} \Delta_{az}^{+ay} \left( \omega_{D44}^z + \omega_{D32}^z \right) - \\ -\omega_{B44}^{y\,2} \omega_{D32}^z \Delta_{ay}^- - \omega_{B44}^{z\,2} \omega_{D44}^z \Delta_{az}^+ \end{bmatrix} \end{aligned} \right\}^{-1} \quad (3)$$

The magnetoelectric susceptibility tensor $\kappa_{ij}^{me}$ has only one non-zero component: $\kappa_{xz}^{me} = -\alpha_{zz} \Delta_{mx}^{pz} \Delta_{mx}^{-1}$.

Here the following notation has been introduced:
$\Delta_{ax,y,z}^{(+,-)} = \omega_{B66,22,32}^{y\,2} \omega_{tx,l,l}^2 - \left( \omega^2 - \omega_{ax,y,z}^{(+,-)\,2} \right) \left( \omega^2 - \omega_{tx,l,l}^2 \right)$,

$\omega_{B66,44}^{y\,2} = \left( 2\mu c_{66,44} \right)^{-1} B_{66,44}^2 A_1^2$,

$\omega_{B44}^{zy\,2} = \omega_{B44}^z \omega_{B44}^y$, $\omega_{B22,32}^{y,z\,2} = 2 \left( \mu c_{22} \right)^{-1} B_{22,32}^2 A_{1,2}^2$,

$\omega_{D44}^{z\,2} = \left( \lambda c_{44} \right)^{-1} D_{44}^2 P_0^2$, $\omega_{B44,66}^{z\,2} = \left( 2\mu c_{44,66} \right)^{-1} B_{44,66}^2 A_2^2$,

$\omega_{D32}^{z\,2} = 4 \left( \lambda c_{22} \right)^{-1} D_{32}^2 P_0^2$,

$\left( \Delta_{az}^+, \Delta_{ay}^- \right) = \omega_{B44}^{y\,2} \omega_{tz}^2 - \left( \omega^2 - \left( \omega_{az}^{+\,2}, \omega_{ay}^{-\,2} \right) \right) \left( \omega^2 - \omega_{tz}^2 \right)$,

$\Delta_{ay,z}^{+az,y} = i\Omega_{-,+}^2 \left( \omega^2 - \omega_{l,tz}^2 \right) + i\omega_{B22,44}^y \omega_{B32,44}^z \omega_{l,tz}^2$,

$\Delta_{py,z} = \omega_{D44,32}^{z\,2} \omega_{tz,l}^2 - \left( \omega^2 - \omega_{py,z}^2 \right) \left( \omega^2 - \omega_{tz,l}^2 \right)$,

$\Delta_{mx} = -\left( \omega^2 - \omega_l^2 \right) \left[ \omega_{2xy}^4 \Delta_{az}^- + \omega_{2xz}^4 \Delta_{ay}^+ + i2\omega_{2xy}^2 \omega_{2xz}^2 \Delta_{ay}^{+az} \right] -$
$- \left( \omega^2 - \omega_{mx}^2 \right) \left( \Delta_{ay}^+ \Delta_{az}^- + \left[ \Delta_{ay}^{+az} \right]^2 \right)$,

$\Delta_{mx}^{pz} = \sqrt{\lambda/\tilde{\mu}} \left\{ \begin{aligned} & \left( \omega^2 - \omega_l^2 \right) \left( \omega_{me}^{Ay2} \Delta_{ay}^{+az} + \omega_{me}^{Az2} \Delta_{az}^- \right) \left( \omega_{2xy}^2 - i\omega_{2xz}^2 \right) - \\ & -i\omega_{D32}^z \omega_l^2 \left( \omega_{B32}^z \Delta_{ay}^{+az} + \omega_{B22}^y \Delta_{az}^- \right) \left( \omega_{2xy}^2 + i\omega_{2xz}^2 \right) \end{aligned} \right\}$,

$\Delta_{pz}^{mx} = \sqrt{\tilde{\mu}/\lambda} \left( \omega^2 - \omega_l^2 \right) \times$
$\times \left\{ \begin{aligned} & \left( \omega^2 - \omega_l^2 \right) \begin{bmatrix} \omega_{2xy}^2 \left( \omega_{me}^{Az2} \Delta_{ay}^{+az} + i\omega_{me}^{Ay2} \Delta_{ay}^+ \right) + \\ + \omega_{2xz}^2 \left( \omega_{me}^{Ay2} \Delta_{ay}^{+az} - i\omega_{me}^{Az2} \Delta_{az}^- \right) \end{bmatrix} - \\ & -\omega_{D32}^z \omega_l^2 \begin{bmatrix} \omega_{2xy}^2 \left( \omega_{B22}^y \Delta_{ay}^{+az} - i\omega_{B32}^z \Delta_{ay}^+ \right) - \\ -\omega_{2xz}^2 \left( \omega_{B32}^z \Delta_{ay}^{+az} + i\omega_{B22}^y \Delta_{az}^- \right) \end{bmatrix} \end{aligned} \right\}$,

$\omega_{ij(y,z)}^2 = \left( \mu \tilde{\mu} \right)^{-1/2} \lambda_i M_{0j} A_{(1,2)}$, $i = 1, 2$, $j = x, y, z$, is the characteristic frequency of exchange interaction between magnetic and antiferromagnetic subsystems, $\omega_{me}^{A(y,z)2} = 2\mu^{-1}\nu A_{(1,2)}k$, $\omega_{tx,z} = s_{tx,z}q$, $\omega_l = s_l q$,

$$\Omega_\pm^2 = \mu^{-1}\left\{\left[u - B_{44}^2(2c_{44})^{-1}\right]A_1 A_2 \pm 2\nu P_0 k\right\}, \quad s_l = \sqrt{c_{22}/\rho},$$

$s_{tx,z} = \sqrt{2c_{66,44}/\rho}$, are the velocities of longitudinal, transversal $x$ and $z$ polarized acoustic waves, $q$ is the wave number of the propagating wave, $\omega_{pi}^2 = \lambda^{-1}\left(b - 8(\mathbf{D_i}, \mathbf{u^{0p}})\right)$ is the characteristic frequency of electro-dipole oscillations, $\mathbf{u^{0p}} = \Delta^{-1}\left(\Delta_{D3}^1, \Delta_{D3}^2, \Delta_{D3}^3\right)$, $\omega_{mi}^2 = \tilde{\mu}^{-1}\beta_i$, $i = x, y, z$, is the characteristic frequency of magnetization oscillations,

$$\omega_{a(x,y,z)}^{\pm^2} =$$
$$= \frac{1}{\mu}\begin{bmatrix} a + \lambda_1 M_{0(x,y,z)} + \lambda_2 \mathbf{M}^2 + 4\left(\mathbf{B_{(1,2,3)}}, \mathbf{u^{0P}}\right) + \\ +\gamma k^2 + \alpha k^4 + \begin{pmatrix} 4^{-1}u\{A_1^2 + A_2^2\} + \\ +2\left(\mathbf{B_{(1,2,3)}}, \mathbf{u^{0A1}} + \mathbf{u^{0A2}}\right)\end{pmatrix} \pm \\ \pm\left(4^{-1}u\{A_1^2 - A_2^2\} + \left(\mathbf{B_{(1,2,3)}}, \mathbf{u^{0A1}} - \mathbf{u^{0A2}}\right)\right) \end{bmatrix}$$

is the characteristic frequency of antiferromagnetic oscillations. $\omega_{Bij}^\alpha$ and $\omega_{Dij}^\alpha$ are the characteristic frequencies of magneto-acoustic and electro-acoustic interaction. $\omega_{me}^{A(y,z)}$ caused by electromagnons contribution into oscillations. $\Omega_\pm^2$ also contains the contribution of electromagnons $2\nu P_0 k\mu^{-1}$ which appear only in modulated ($k \neq 0$) non-collinear phase (in collinear phase $P_0 = 0$).

Equations $\Delta_{ai}^{(+,-)} = 0$ ($i = x,y,z$) correspond to the spectrum of coupled antiferromagnetic $i$-polarized and acoustic waves. Equations like $\Delta_{ay}^+ \Delta_{az}^- + \left[\Delta_{ay}^{+az}\right]^2 = 0$ correspond to the spectrum of coupled antiferromagnetic, acoustic and electro-dipole waves caused by magnetostriction and magnetoelectricity. Equations like $\Delta_{mx}\Delta_{pz} - \Delta_{mx}^{pz}\Delta_{pz}^{mx} = 0$ correspond to spectrum of coupled antiferromagnetic, magnetic, electro-dipole and acoustic modes of oscillations caused by magnetostriction, electrostriction, magnetoelectricity and exchange interactions between antiferromagnetic and ferromagnetic subsystems. If we put the denominators of $\chi_{yy}$ and $\chi_{zz}$ equal to zero, we will get the spectrums of coupled ferro- and antiferromagnetic oscillations.

When $\mathbf{H} = 0$, $\kappa_{ij}^{me} = 0$. So, an external magnetic field activates some components of the tensor. Note, that this is the feature of the model. In ground state we put $\mathbf{M_0} = 0$ in absence of external magnetic field. Indeed the magnetization is very small, but not equal to zero.

Since $k$, $A_1$, $A_2$, $P_0$ and $M_{0i}$ are the functions of $\mathbf{H}$ and $E_z$, defined from solution (2), so the resonant frequency values depend on values of both electric and magnetic fields. For values of constants (TbMnO$_3$) $\gamma \sim -10^{-14}$ cm$^2$, $\alpha \sim 10^{-28}$ cm$^4$, $a \sim 100$, $u \sim 0.1$, $w \sim 10$, $\beta \sim 100$, $b \sim 0.4$, $\nu \sim 10^{-9}$, $\lambda_{1,2} \sim 10^{-4}$, for example, $\omega_{1xz}^2$ vary in range of $0 \leq \omega_{1xz}^2 \leq \omega_{1xz\max}^2$ with changing of $H_x$, and take a maximum about $\omega_{1xz\max}^2 \sim 10^{21}$ s$^{-2}$ in field $H_x \sim 10^5$ Oe (10 T). Fig. 1 shows this dependence. The frequency $\Omega_\pm$ changes with the electric field on value about $\Delta\Omega \approx \left(\nu k\mu^{-1}b^{-1}E_z\right)^{1/2} \sim 10^{10}$ s$^{-1}$ for $E_z \sim 100$ CGSE (about 30 kV/cm). It shifts the resonant frequency corresponding to interaction of electromagnetic wave with antiferromagnetic subsystem to lower frequencies on value about $\omega_{1xz}$ and to higher frequencies on $\Delta\Omega$.

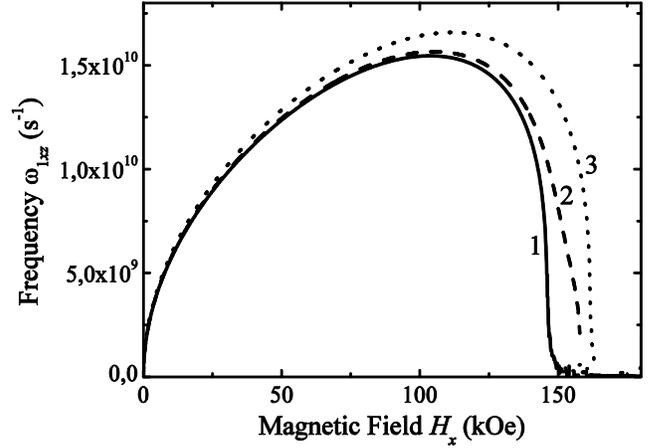

Fig. 1. Field dependence of shift frequency $\omega_{1xz}$ of orthorhombic perovskite multiferroic (TbMnO$_3$) placed in magnetic field $\mathbf{H}\|\mathbf{x}$ and electric field $\mathbf{E}\|\mathbf{z}$. 1 – $E = 0$ CGSE, 2 – $E = 100$ CGSE (about 30 kV/cm), 3 – $E = 500$ CGSE (about 150 kV/cm). 10 kOe = 1 T.

## IV. SPECTRUM OF COUPLED WAVES

To investigate the spectrum of coupled waves in orthorhombic multiferroic one have to take into account the system of Maxwell's equations with material equations $p_i = \alpha_{ij}e_j + \kappa_{ij}^{em}h_j$, $m_i = \chi_{ij}h_j + \kappa_{ij}^{me}e_j$, where $\alpha_{ij}$, $\chi_{ij}$ and, $\kappa_{ij}^{me}$ are given in [11]. Study shows that dispersion equations are the followed:

$$\mathbf{H}\|\mathbf{x}: \quad c^2q^2\mu_{yy} = \omega^2\varepsilon_{xx}\mu_{zz}, \quad \mathbf{e}\|\mathbf{x}$$
$$\left(cq/\omega - 4\pi\kappa_{zx}^{em}\right)\left(cq/\omega - 4\pi\kappa_{xz}^{me}\right)\mu_{yy} = \varepsilon_{zz}\mu_{xx}, \quad \mathbf{e}\|\mathbf{z}$$
$$\mathbf{H}\|\mathbf{y}: \quad c^2q^2\mu_{yy} = \omega^2\left(\varepsilon_{zz}\mu_{zz} - 16\pi^2\left|\kappa_{yz}^{me}\right|^2\right), \quad \mathbf{e}\|\mathbf{z}$$
$$c^2q^2\varepsilon_{yy} = \omega^2\varepsilon_{xx}\left(\varepsilon_{yy}\mu_{zz} - 16\pi^2\left|\kappa_{zy}^{me}\right|^2\right), \quad \mathbf{e} = (e_x, e_y, 0) \quad (4)$$
$$\mathbf{H}\|\mathbf{z}: \quad \varepsilon_{yy}\mu_{yy} - 16\pi^2\left|\kappa_{zz}^{me}\right|^2 = 0, \quad \mathbf{e}\|\mathbf{y}$$
$$\left(\mu_{yy}c^2q^2/\omega^2\mu_{xx} - \varepsilon_{zz}\mu_{zz}\right) \times$$
$$\times\left(c^2q^2/\omega^2\varepsilon_{xx} - \mu_{zz}\right) = 16\pi^2\varepsilon_{xx}\left|\kappa_{zz}^{me}\right|^2, \quad \mathbf{e} = (e_x, 0, e_z).$$

Here, $\varepsilon_{ii} = 1 + 4\pi\alpha_{ii}$, $\mu_{ii} = 1 + 4\pi\chi_{ii}$ – permittivity and permeability tensors, consequently, c – velocity of speed in vacuum.

So, wave will split into two waves of different polarizations. In some cases, transversal electromagnetic wave will excite a longitudinal one, and *vice versa*. Solving the dispersion equations (4), we will get a spectrum of the coupled spin, acoustic, electro-dipole and electromagnetic waves.

Fig. 2 shows the spectrum of coupled waves in case of magnetic field directed along *y*-axis with *z*-polarized electric component of electromagnetic wave. The spectrum has a complex band structure. Band gaps observed as for electromagnetic, such for acoustic waves. This effect is manifested due to resonant interaction between subsystems of multiferroic. Width and frequencies of band gaps depend on external electric and magnetic field values. On the lower inset is shown the interaction of ferromagnetic and acoustic modes of oscillations. One can see that increasing of magnetic field leads to increasing of frequency of magnetic oscillations and increasing of magneto-acoustic gap. When magnetic field is more than $H_c$ (about 75 kOe or 7.5 T in case of magnetic field directed along y-axis for TbMnO$_3$ [1]), antiferromagnetic structure transforms from the non-collinear cycloidal to the collinear sinusoidal ($A_1 = 0$) and the spontaneous polarization along z-axis disappears, and the magnetoelectric mode suppress [15]. Applying of magnetic field a little more than the critical value leads to abruptly decrease of frequency of magnetic oscillations and magneto-acoustic gap. Applied external electric field induces polarization, and leads to resuming of cycloidal structure. So, the value of critical magnetic field shifts with applying of an external electric field to higher values. It leads to decreasing of magneto-acoustic gap and increasing of magnetic oscillations frequency by electric field. On the top inset is shown the interaction of antiferromagnetic and electromagnetic modes. Applying of external electric field leads to increasing of antiferromagnetic oscillation frequency. As mentioned above, in absence of external magnetic field $\kappa_{ij}^{me} = 0$ and the alternating electric field will no excite magnetic oscillations. How it has been shown in [11], $\kappa_{ij}^{me}$ significantly differ from zero only near the resonant. And exciting of longitudinal component of electromagnetic wave will be observed only in this narrow frequency range.

## V. Conclusion

Study on spectrum of coupled spin, acoustic, electro-dipole and electromagnetic waves in orthorhombic perovskite multiferroics showed that the spectrum has a complex band structure. Its width and frequency ranges depend on electric and magnetic field values and directions. The electric, magnetic, and magnetoelectric susceptibility tensors depend on external fields as well. Since reflectance and transmittance of electromagnetic waves can be expressed trough susceptibilities, it is possible to tune these characteristics by electric and magnetic fields. Electromagnetic wave of arbitrary polarization decays into two waves of different polarization with different phase velocities. This effect is observed mostly at frequencies near the band gaps. The difference between the velocities depend on external fields values. So, one can control the electrodynamic properties of such multiferroic by both electric and magnetic fields.

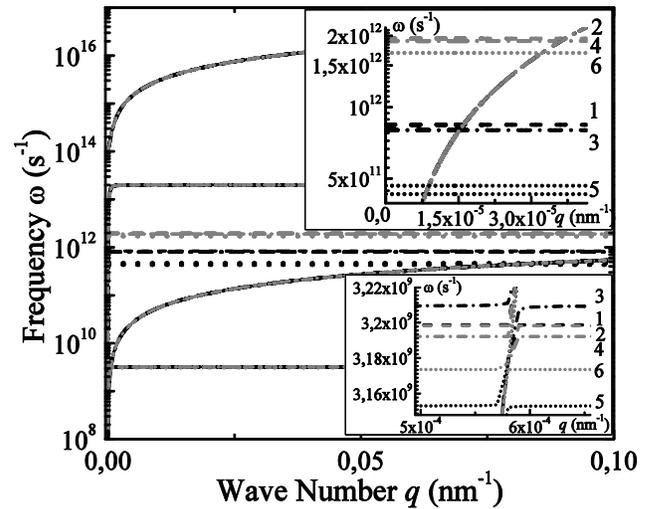

Fig. 2. Spectrum of coupled waves of orthorhombic perovskite multiferroic (TbMnO$_3$) placed in magnetic field **H**||**y** and electric field **E**||**z**. 1 – $H$ = 10 kOe (1 T), $E$ = 0 CGSE, 2 – $H$ = 10 kOe (1 T), $E$ = 100 CGSE (about 30 kV/cm), 3 – $H$ = 40 kOe (4 T), $E$ = 0 CGSE, 4 – $H$ = 40 kOe (4 T), $E$ = 100 CGSE (about 30 kV/cm), 5 – $H$ = 75 kOe (7.5 T), $E$ = 0 CGSE, 6 – $H$ = 75 kOe (7.5 T), $E$ = 100 CGSE (about 30 kV/cm).

## References

[1] T. Kimura, T. Goto, H. Shintani, K. Ishizaka, T. Arima, and Y. Tokura, "Magnetic control of ferroelectric polarization", *Nature (London)*, vol. 426, pp. 55–58, Nov. 2003.
[2] T. Kimura, G. Lawes, T. Goto, Y. Tokura, and A. P. Ramirez, "Magnetoelectric phase diagrams of orthorhombic RMnO$_3$", *Phys. Rev. B*, vol. 71, p. 224425, June 2005.
[3] M. Kenzelmann, A.B. Harris, S. Jonas, C. Broholm, J. Schefer, S.B. Kim, C.L. Zhang, S.-W. Cheong, O.P. Vajk, and J.W. Lynn, "Magnetic Inversion Symmetry Breaking and Ferroelectricity in TbMnO$_3$", *Phys. Rev. Lett.*, vol. 95, p. 087206, Aug. 2005.
[4] Binbin Wu and Gerald J. Diebold, "Photoacoustic effect in a sinusoidally modulated structure", *Appl. Phys. Lett.*, vol. 100, p. 164102, April 2012.
[5] I.V. Bychkov, D.A. Kuzmin, V.G. Shavrov, "Hybridization of electromagnetic, spin and acoustic waves in magnetic having conical spiral ferromagnetic order", *J Magn. Magn. Mater.*, vol. 329, pp. 142-145, March 2013.
[6] I.E. Chupis, "Excitation of electromagnons in the ferroelectromagnet TbMnO$_3$ by an alternating electric field", *Low Temp. Phys.*, vol. 35, p. 858, Nov. 2009.
[7] I.E. Chupis, "Polaritons in the ferromagnetic material TbMnO$_3$ at a boundary with a metal in a magnetic field", *Low Temp. Phys.*, vol. 38, p. 175, Feb. 2012.
[8] R. Valdés Aguilar, M. Mostovoy, A.B. Sushkov, C.L. Zhang, Y.J. Choi, S-W. Cheong, and H.D. Drew, "Origin of Electromagnon Excitations in Multiferroic RMnO$_3$", *Phys. Rev. Lett.*, vol. 102, p. 047203, Jan. 2009.
[9] J. S. Lee, N. Kida, S. Miyahara, Y. Takahashi, Y. Yamasaki, R. Shimano, N. Furukawa and Y. Tokura, "Systematics of electromagnons in the spiral spin-ordered states of RMnO$_3$", *Phys. Rev. B*, vol. 79, p. 180403, May 2009.
[10] A. Pimenov, A.M. Shuvaev, A.A. Mukhin, and A. Loidl, "Electromagnons in multiferroic manganites", *J. Phys.: Condens. Matter*, vol. 20, p. 434209, Oct. 2008
[11] I.V. Bychkov, D.A. Kuzmin, S.J. Lamekhov, and V.G. Shavrov, "Magnetoelectric susceptibility tensor of multiferroic TbMnO$_3$ with cycloidal antiferromagnetic structure in external field", *J. Appl. Phys.*, vol. 113, p. 17C726, Apr. 2013.
[12] T. Kimura, S. Ishihara, H. Shintani, T. Arima, K. T. Takahashi, K. Ishizaka, and Y. Tokura, "Distorted perovskite with $e_g^1$ configuration as a frustrated spin system", *Phys. Rev. B*, vol. 68, p. 60403(R), Aug. 2003.
[13] I.E. Chupis, "Magnetoelectric states of TbMnO$_3$ in magnetic fields of different directions", *Low Temp. Phys.*, vol. 34, p. 422, Jun. 2008.
[14] J.L. Ribeiro and L.G. Vieira, "Landau model for the phase diagrams of the orthorhombic rare-earth manganites RMnO$_3$ (R= Eu, Gd, Tb, Dy, Ho)", *Phys. Rev. B.*, vol. 82, p. 064410, Aug. 2010.
[15] A. Pimenov, A.A. Mukhin, V.Yu. Ivanov, V.D. Travkin, A.M. Balbashov, and A. Loidl, "Possible evidence for electromagnons in multiferroic manganites", *Nat. Phys.*, vol. 2, p. 97, Jan. 2006.